%% file: main.tex
\documentclass[cameraready]{Interspeech}
\usepackage{bm}
\usepackage{booktabs}
\usepackage{multirow}

\title{Do speech foundation models perceive speaker similarity as humans do?}

\author[affiliation={1},equalcontribution]{Minoru}{Kishi}
\author[affiliation={1},equalcontribution]{Hayato}{Yagi}
\author[affiliation={1,2},orcid=0000-0003-0520-7847]{Shinnosuke}{Takamichi}
\author[affiliation={2},orcid=0000-0002-7967-2613]{Yuki}{Saito}

\address{
    $^1$ Keio University, Japan \\
    $^2$ The University of Tokyo, Japan
}

\email{\{minorukishi1091, hayatobuti523, shinnosuke\_takamichi\}@keio.jp, yuuki\_saito@ipc.i.u-tokyo.ac.jp}


\usepackage{comment}

\begin{document}

\maketitle

\begin{abstract} \vspace{-1mm}
    This study presents a comparative analysis between the speaker embeddings of speech foundation models and human subjective perception of speaker similarity. Human listeners have the ability to judge speaker similarity on a continuous scale discerning how similar two voices are. In contrast, speech foundation models embed speaker characteristics into numerical representation. However, a question remains: does the numerical distance between speaker embeddings in these models truly align with the similarity perceived by humans? To address this, we conduct a comprehensive investigation using more than 40 models to compare model-derived distances with human-perceived similarity scores. Furthermore, we identify which factors in model configuration contribute most to a speaker embedding that mirrors human perception. Our findings provide insights for the development of more perceptually grounded speech foundation models.
\end{abstract}

\input{sections/1.introduction}

\input{sections/2.related-work}

\input{sections/3.proposed}

\input{sections/4.experiments}

\input{sections/5.conclusion}

\clearpage

\input{sections/acknowledgments}
\input{sections/disclosure_ai}

\bibliographystyle{IEEEtran}
\bibliography{mybib}

\end{document}

%% file: sections/1.introduction.tex
\vspace{-2mm}
\section{Introduction}\vspace{-1mm}
    Speech carries speaker information, which human listeners can effortlessly perceive~\cite{belin2000}. Beyond simple speaker verification discerning whether two voices belong to the same speaker, humans possess the ability to judge the degree of similarity or dissimilarity between different speakers~\cite{kitamura2015measurement}. This \textit{perceptual speaker similarity}~\cite{saito2021} represents a high-level cognitive function that cannot be fully explained by acoustic features alone~\cite{liu2024}.

    Recent speech foundation models (self-supervised models (SSLs)~\cite{wav2vec2, hubert, wavlm} and large supervised learning models~\cite{whisper, parakeet}) have demonstrated a remarkable capacity to capture speaker characteristics. The fact that their internal representations are highly effective for speaker verification~\cite{9747814, wavlmsslsv}, as evidenced by various probing experiments~\cite{chiu2025largescaleprobinganalysisspeakerspecific, ashihara2024}, confirms that these models acquire discriminative speaker embeddings.
    
    However, a question remains: \textit{do these speech foundation models capture the ``degree of similarity'' between speakers in a manner consistent with human perception?} In other words, how does the geometric relationship between speaker embeddings (i.e., \textit{speaker-embedding similarity}) correspond to the perceptual similarity? If these models have learned representations that align with human perception, it would suggest that high-level perceptual function can emerge solely from data. Furthermore, identifying which model configurations contribute to this alignment would provide critical guidance for developing speech foundation models with human-like speaker perception.

\begin{figure}[t]
      \centering
      \includegraphics[width=0.98\linewidth]{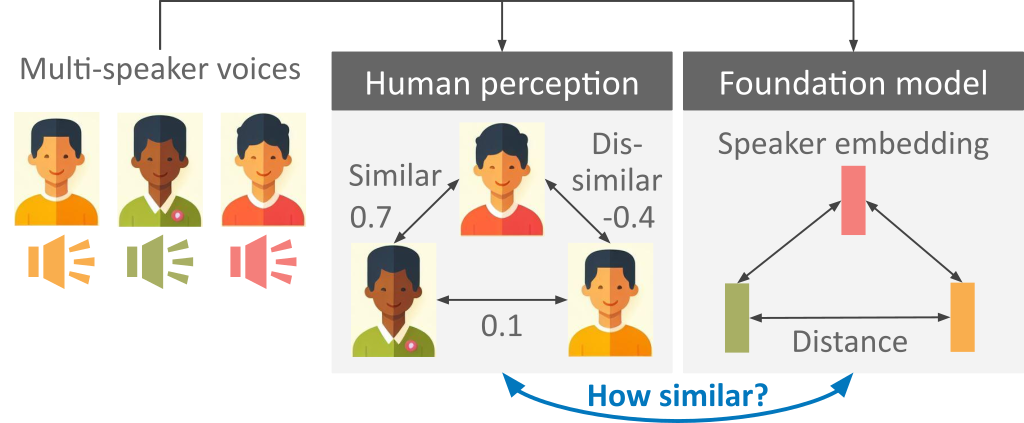} 
      \vspace{-2mm}
      \caption{Concept of this paper.}
      \vspace{-3mm}
      \label{fig:concept}
    \end{figure}

    To address these questions, this study investigates the correspondence between speaker-embedding similarity and perceptual similarity (see Fig.~\ref{fig:concept}). Utilizing over $40$ models and perceptual similarity scores~\cite{jvs, cstr-vctk}, we analyze which models and layers yield embeddings that align with human perception. Additionally, we perform a multiple regression analysis to quantify how various model configurations influence this alignment.

%% file: sections/2.related-work.tex
\vspace{-2mm}
\section{Related work} \vspace{-1mm}
    \subsection{Perceptual speaker similarity} \vspace{-2mm}
        The perceptual speaker similarity score is derived by presenting pairs of voices from two different speakers to human listeners. The listeners were asked to quantify the degree of similarity. For instance, a previous work~\cite{saito2021} asked multiple listeners to provide ratings on a $7$-point scale ranging from $-3$ (very dissimilar) to $+3$ (very similar). Obtained scores for a given speaker pair are averaged across all listeners. By repeating this process for all possible speaker combinations within a speaker set, the collected data can be represented 
        as a weighted undirected graph (nodes are individual speakers, edge weights are the similarity scores).
    
    
        
        

    \vspace{-1mm}
    \subsection{Analysis of speech foundation models} \vspace{-2mm}
        \label{sec:2.2}

        Most speech foundation models are composed of a series of stacked Transformer layers~\cite{hubert, wav2vec2, wavlm}, and extensive studies have been conducted to characterize the specific features captured by each layer. Lower layers primarily capture low-level acoustic information, such as pitch, energy~\cite{chiu2025largescaleprobinganalysisspeakerspecific}, and gender~\cite{Krishnan2024}. In contrast, middle layers exhibit strong performance in speaker identification (SID) and automatic speaker verification (ASV)~\cite{ashihara2024}, suggesting that discriminative speaker-related information is most prominently represented in these intermediate stages.

        In this study, we evaluate our findings regarding perceptual speaker similarity in light of these established layer-wise characteristics. In experiments, we discuss the results by aligning the findings with the aforementioned results on probing.

%% file: sections/3.proposed.tex
\vspace{-2mm}
\section{Methodology} \vspace{-1mm}
    \subsection{Perceptual similarity} \vspace{-2mm}
        Let $S$ denote a set of speakers. For any speaker pair $(i, j) \in S$ where $i \neq j$, we define the perceptual similarity score as $c_{i,j}^{\text{(human)}} \in \mathbb{R}$. The complete set of these scores is represented as $W^{\text{(human)}} = \{ c_{i,j}^{\text{(human)}} \mid i, j \in S, i \neq j \}$. 
        
        When represented as a weighted undirected graph, we denote it as $G^{\text{(human)}} = \left(S, E, W^{\text{(human)}}\right)$. The set of nodes and set of edge weights are equivalent to $S$, and $W^{\text{(human)}}$, respectively. $E = \{ \{i, j\} \mid i, j \in S, i \neq j \}$ is the set of edges. This graph is considered a complete graph (excluding self-loops), as edges are defined for all possible pairs of distinct speakers in $S$.

    \vspace{-1mm}
    \subsection{Speaker-embedding similarity}\vspace{-2mm}
        To extract speaker embeddings, we define $f(\cdot)$ as the transformation function of a given speech foundation model from the input layer to a specific Transformer layer. When an utterance $x_i$ from speaker $i$ is fed into the model, the output of the Transformer layer is represented as a sequence of hidden states $f(x_i) = \{ \bm{h}_{i, t} \}_{t=1}^{T}$, where $T$ denotes the number of frames. The speaker embedding $\bm{e}_i$ is then obtained by calculating the mean vector of these hidden states across all frames and all available utterances for that speaker. This process is repeated for all speakers in the set $S$.

        The resulting representation is modeled as a weighted undirected graph $G^{\text{(model)}} = (S, E, W^{\text{(model)}})$. The similarity score $c_{i,j}^{\text{(model)}}$ for a speaker pair $(i, j)$ is defined as the cosine similarity between their respective embeddings $\bm{e}_i$ and $\bm{e}_j$. The set of edge weights is $W^{\text{(model)}} = \{ c_{i,j}^{\text{(model)}} \mid i, j \in S, i \neq j \}$.

    \vspace{-1mm}
    \subsection{Correspondence between perceptual speaker similarity and speaker-embedding similarity} \label{sec:3.3} \vspace{-2mm}
        To quantify the alignment between human perception and model-derived representations, we evaluate the correspondence between the two previously defined graphs, $G^{\text{(human)}}$ and $G^{\text{(model)}}$, using the following three metrics:
    

        \textbf{Pairwise similarity correlation.}
        We first evaluate the direct correlation between the edge weights of both graphs. Specifically, we calculate the Pearson (LCC) and Spearman correlation coefficients (SRCC) between the sets of scores $\{c_{i,j}^{\text{(human)}}\}$ and $\{c_{i,j}^{\text{(model)}}\}$. Since the graphs are undirected, we deduplicate symmetric pairs (e.g., $c_{i,j}$ and $c_{j,i}$) to ensure each unique speaker pair is represented only once in the correlation analysis.

        \textbf{Frobenius distance.}
        We treat the sets of similarity scores as adjacency matrices (similarity matrices for weighted graph) to measure the structural distance between the two representations. Let $\mathbf{A}^{\text{(human)}}$ and $\mathbf{A}^{\text{(model)}}$ denote the adjacency matrices where the $(i, j)$-th elements are $c_{i,j}^{\text{(human)}}$ and $c_{i,j}^{\text{(model)}}$, respectively. We then compute the Frobenius distance between these two matrices to quantify their element-wise discrepancy.
        
        \textbf{Spectral distance.}
        To compare the global topological structures of the graphs, we employ spectral analysis. For each graph's adjacency matrix, $\mathbf{A}^{\text{(human)}}$ or $\mathbf{A}^{\text{(model)}}$, we compute the normalized graph Laplacian and its corresponding eigenvalues. We then calculate the $\ell_2$ distance between the $k$ smallest non-zero eigenvalues. This approach effectively measures the similarity in the ``low-graph-frequency'' components, reflecting the consistency of the global speaker-clustering structure between human perception and the model's embedding.




%% file: sections/4.experiments.tex
\vspace{-2mm}
\section{Experiments} \vspace{-1mm}
    \subsection{Experimental settings} \vspace{-2mm}
    
        \textbf{Perceptual similarity.}
        We used two speech datasets containing perceptual similarity scores: JVS ($49$ males and $51$ females)~\cite{jvs} and VCTK ($52$ females)\footnote{Scores on VCTK male speakers are not publicly available.}~\cite{cstr-vctk,vctk-sim}. In each dataset, perceptual similarity scores in the range of $[-3, 3]$ are provided for all intra-gender speaker combinations. For our experiments, these scores were normalized to the range of $[0, 1]$. 

        \textbf{Speaker-embedding similarity.}
        We used a total of $43$ open-source models, categorized into the following six types. Labels we used follow official names of models.  
        \begin{itemize}
            \item \textbf{Supervised ASR (automatic speech recognition) models}: 
            Parakeet (\texttt{parakeet-\{ctc, rnnt\}-\{0.6b, 1.1b\}}, 
            \texttt{parakeet-tdt-0.6b-\{v2, v3\}}, 
            \texttt{parakeet-tdt-1.1b})~\cite{parakeet}\footnote{\label{ftn:nvidia-model}\url{https://huggingface.co/nvidia/models}} and 
            Whisper (\texttt{whisper-\{small, base, medium, large\}})~\cite{whisper}\footnote{\url{https://github.com/openai/whisper}}) 
        
            \item \textbf{Supervised TTS (text-to-speech) models}: 
            Qwen3-TTS (\texttt{qwen3-tts-12hz-\{0.6b, 1.7b\}-base})~\cite{Qwen3-TTS}\footnote{\url{https://github.com/QwenLM/Qwen3-TTS}}, 
            SpeechT5 (\texttt{speecht5-\{base, large, tts\}})~\cite{speechT5}\footnote{\url{https://github.com/microsoft/SpeechT5}, \url{https://huggingface.co/microsoft/speecht5_tts}}, 
            SpeechGPT (\texttt{speechgpt-7b-\{ma, cm\}})~\cite{speechgpt}\footnote{\url{https://github.com/0nutation/SpeechGPT}}, and 
            VALL-E X (\texttt{vallex-\{q01--q08\}})~\cite{vallex}\footnote{\url{https://github.com/Plachtaa/VALL-E-X} (\texttt{\{q01--q08\}} indicates index of the codebook.)}. 
        
            \item \textbf{Supervised TTA (text-to-audio) models}: 
            AudioGen (\texttt{audiogen-medium})~\cite{audiogen}\footnote{\label{ftn:facebook-model}\url{https://huggingface.co/facebook/models}}. 
        
            \item \textbf{Supervised audio classification models:} 
            AST (\texttt{ast\_audioset\_10\_10\_0.4593})~\cite{ast}\footnote{\url{https://github.com/YuanGongND/ast}}. 
        
            \item \textbf{Speech SSL models}: 
            HuBERT (\texttt{hubert-\{base-ls960, large-ll60k, large-ls960-ft\}})~\cite{hubert}\footref{ftn:facebook-model}, 
            wav2vec 2.0 (\texttt{wav2vec2-\{base, base-960h, large, large-960h, large-xlsr-53\}})~\cite{wav2vec2}\footref{ftn:facebook-model}, and 
            WavLM (\texttt{wavlm-\{base, base+, large, ssl\_sv\}})~\cite{wavlm,wavlmsslsv}\footnote{\url{https://huggingface.co/docs/transformers/model_doc/wavlm,https://github.com/theolepage/wavlm_ssl_sv/blob/main/README.md}}. 
        
            \item \textbf{Audio SSL models}: 
            ATST-Frame (\texttt{atstframe\_base})~\cite{atstframe}\footnote{\url{https://github.com/Audio-WestlakeU/audiossl/blob/main/audiossl/methods/ATST-Frame/README.md}}. 
        \end{itemize}
        Audio-related models (TTA, audio classification, and audio SSL models) were included to investigate the impact of training on general audio data rather than speech-only data. For TTS/TTA models, we selected architectures capable of taking speech/audio prompts as input. During feature extraction, only the speech prompt was provided, while other inputs (e.g., text) were treated as padding to isolate the model's internal speaker representations. For fair comparison, speaker embeddings were uniformly extracted from $10$ utterances for each speaker.
        Each embedding dimension was standardized using z-score normalization. Speaker-embedding similarity was normalized to the range $[0,1]$.

    \textbf{Spectral distance setting.}
    In the spectral analysis described in Section~\ref{sec:3.3}, we set the number of eigenvalues to $k=10$.
    We confirmed that varying $k$ within a reasonable range does not significantly change the overall trends of the results.

    \vspace{-1mm}
    \subsection{Overall trends: qualitative analysis} \vspace{-2mm}
    \label{sec:4.2}
        \begin{figure*}[t]
          \centering
          \includegraphics[width=0.98\linewidth]{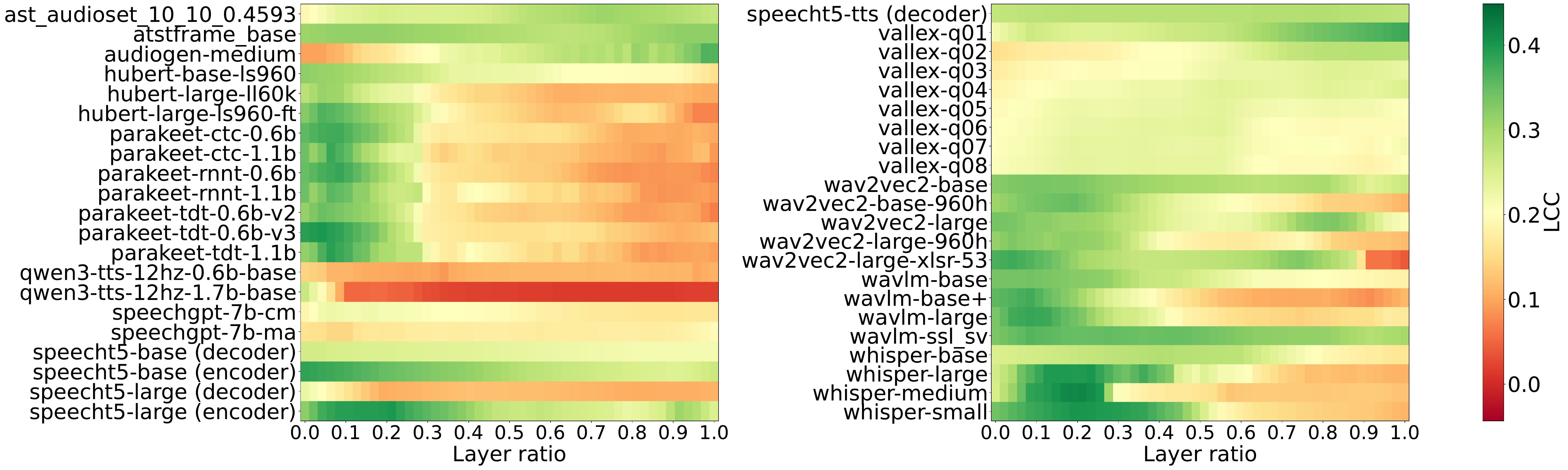} 
          \vspace{-2mm}
          \caption{LCCs of all models and Transformer layers.}
          \vspace{-2mm}
          \label{fig:overall}
        \end{figure*}
    
        To capture the overall trends, we first calculated the correspondence scores for all Transformer layers across all evaluated models. Figure~\ref{fig:overall} shows the LCC averaged across all datasets. Since the number of Transformer layers varies by model, the layer indices are normalized to the range of $[0, 1]$. From this visualization, we can derive the following observations. Although omitted for brevity, similar trends were observed across other datasets and evaluation metrics.

        \textbf{Substantial variance in scores across models.}
        For instance, while WavLM consistently exhibits high scores across most layers, Qwen3-TTS shows relatively lower correspondence. This suggests that the correspondence is heavily influenced by the specific model configuration.

        \textbf{Diverse trends in score progression across layers.}
        We observed distinct patterns in how similarity correspondence changes with layer depth. In many models, the correlation tends to degrade in deeper layers. However, some models, such as AudioGen and VALL-E X, exhibit an upward trend. These results indicate that the slope of the score progression is a characteristic determined by the model configuration.

    \vspace{-1mm}
    \subsection{Effects of model configuration: multiple regression analysis} \vspace{-2mm}
        \input{tables/mean}

        We performed multiple regression analysis to clarify which model configurations contribute to the overall trends observed in Section~\ref{sec:4.2}. The explanatory variables are defined as follows:
        \begin{itemize}
            \item \texttt{is\_dec}: A binary variable indicating whether the decoder component is used for feature extraction ($1$) or the encoder component is used ($0$).
            \item \texttt{is\_mlang}: A binary variable indicating whether the model was trained on multilingual ($1$) or monolingual ($0$) data.
            \item \texttt{is\_ssl}: A binary variable indicating self-supervised learning models ($1$) or supervised learning models ($0$). Fine-tuned SSL models were treated as supervised (0). 
            \item \texttt{hours}: The size of the training data [hours] in a log scale.
            \item \texttt{params}: The number of model parameters in a log scale.
        \end{itemize}
        The binary variables represent categorical configurations, while the continuous variables (\texttt{hours}, \texttt{params}) are standardized. We define two response variables to characterize the layer-wise behavior of each model:
        \begin{itemize}
            \item \texttt{layer\_max}: The max metric value across all Transformer layers in the model.
            \item \texttt{layer\_slope}: The slope obtained by linear regression of the metric values against the normalized Transformer layer indices ($[0, 1]$).
        \end{itemize}  
        These variables were calculated for the four metrics described in Section~\ref{sec:3.3}. For Frobenius and spectral distances, negative values were used so that ``higher is better,'' consistent with LCC and SRCC. To ensure the reliability of the analysis, preliminary diagnostics showed that the Variance Inflation Factor (VIF) for all variables was sufficiently low, indicating that multicollinearity did not affect the results. Furthermore, to isolate the effects of model configurations, we incorporated dataset fixed effects into the regression model to control for potential biases arising from inherent differences between datasets (e.g., JVS and VCTK). Note that the audio-related models were excluded from this regression to focus on speech models. Table~\ref{tab:mean} summarizes the regression results. From these results, we derive the following insights:

        \textbf{Encoder-based architectures foster human-like speaker representations.}
        The coefficient for \texttt{is\_dec} is significantly negative ($p < 0.001$) across three metrics for \texttt{layer\_max}. This indicates that decoder-based architectures tend to form speaker-embedding similarities that align less with human perceptual similarity compared to encoder-based architectures.

        \textbf{Large-scale supervised models align more closely with human perception.}
        Similarly, \texttt{is\_ssl} shows a significant negative correlation ($p < 0.02$) with \texttt{layer\_max} across three metrics. This suggests that large-scale supervised models are more effective at learning speaker representations that correspond to human perception than their SSL counterparts.

        \textbf{Increased parameter scale lowers average alignment but flattens the layer-wise progression.}
        For LCC and SRCC, \texttt{params} has a negative effect on \texttt{layer\_max} but a positive effect on \texttt{layer\_slope}. As the general trend for \texttt{layer\_slope} is negative (as discussed in Section~\ref{sec:4.2}), a positive coefficient here implies that increasing the model size leads to a flatter score distribution across layers. A similar observation can be made for the \texttt{is\_dec} variable.

        \textbf{Model configuration strongly explains peak scores, but less so for layer-wise trends.}
        The coefficient of determination ($R^2$) reaches about $0.8$ for \texttt{layer\_max}, whereas it remains around $0.2$ for \texttt{layer\_slope}. This discrepancy suggests that while the overall quality of speaker representations is well-explained by the current model configurations, the relative contribution of specific layers to this emergence is influenced by factors not captured by our current explanatory variables.

    \vspace{-1mm}
    \subsection{Findings on individual models/factors} \vspace{-1mm}

        \subsubsection{Effect of fine-tuning SSL models} \vspace{-1mm}

\begin{figure}[t]
            \centering
            \includegraphics[width=\linewidth]{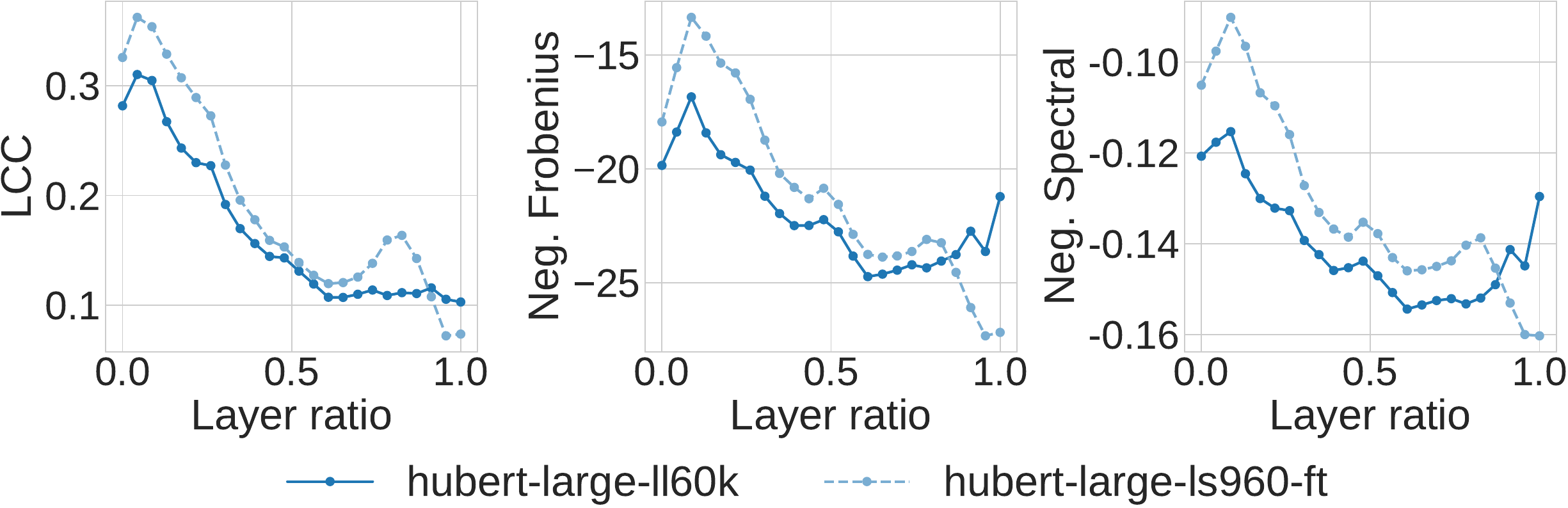}
            \includegraphics[width=\linewidth]{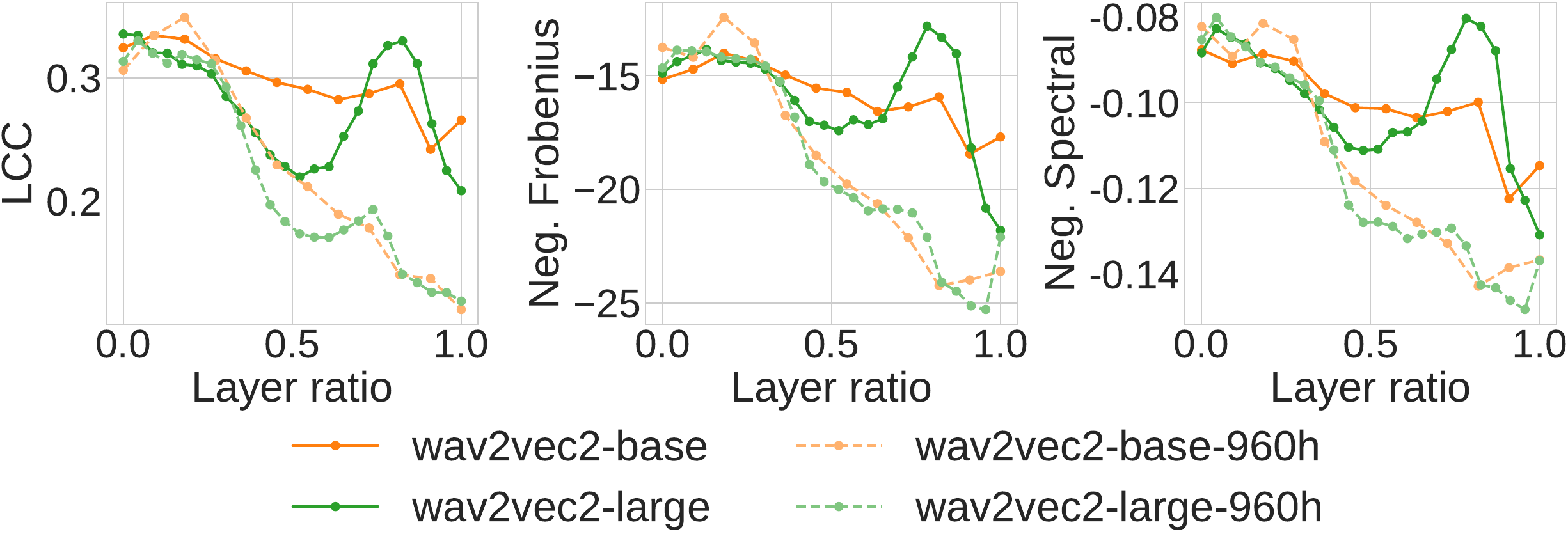}
            \includegraphics[width=\linewidth]{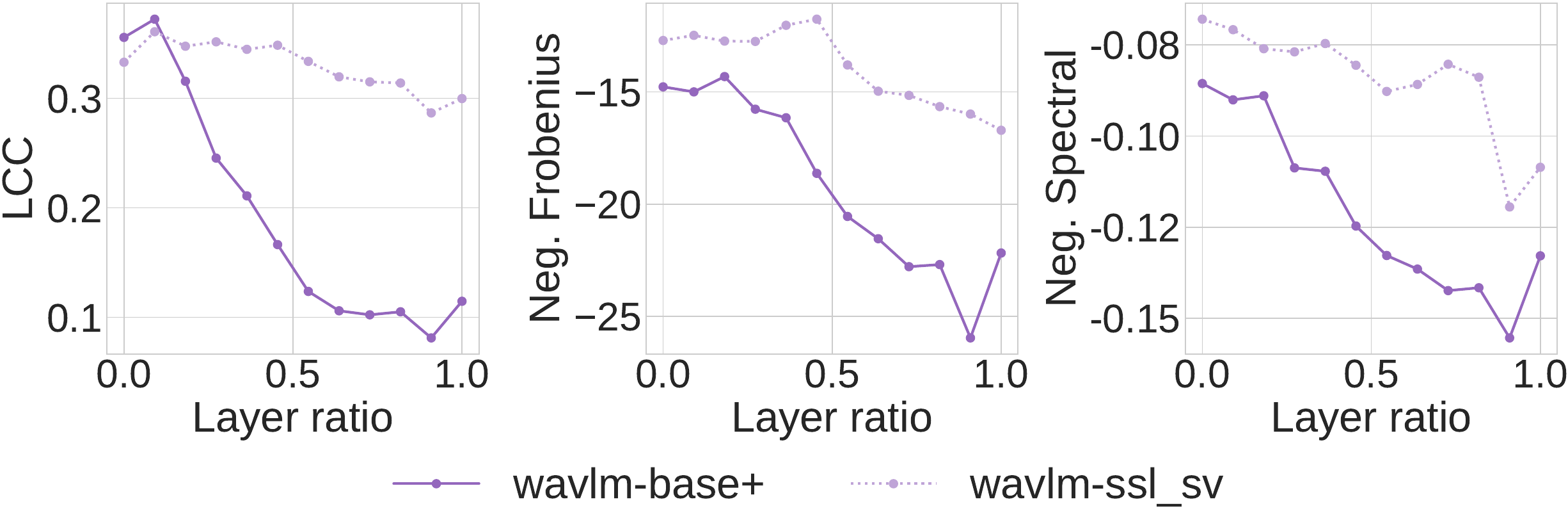}
            \caption{Layer-wise trends for SSL models.}
            \label{fig:lcc_frob_all}
        \end{figure}

        A clear difference emerges depending on the fine-tuning objective of SSL models. 
        In models fine-tuned for ASR, such as \texttt{hubert-large-ls960-ft} and \texttt{wav2vec2-\{base-960h, large-960h\}}, compared with their SSL-only counterparts \texttt{hubert-large-ll60k} and \texttt{wav2vec2-\{base, large\}}, the alignment between model speaker-embedding similarity and human perceptual similarity deteriorates toward deeper layers.
        This is likely because ASR fine-tuning shifts representations toward linguistic content and suppresses speaker-related variations~\cite{pased2021}. As shown in the top and middle rows of Fig.~\ref{fig:lcc_frob_all}, the ASR fine-tuned models exhibit a decrease in all metrics.
        In contrast, models optimized for speaker-related tasks show a different trend. 
        As shown in the bottom row of Fig.~\ref{fig:lcc_frob_all}, compared with the \texttt{wavlm-base+} (SSL only), \texttt{wavlm-ssl\_sv} proposed for speaker representation learning~\cite{wavlmsslsv} exhibits a much flatter layer-wise trend.
        In this case, all metrics remain relatively stable across layers, indicating that the speaker similarity structure is preserved throughout the network. 

        These observations suggest that the fine-tuning objective affects the layer-wise organization of representations in SSL models, consistent with prior studies showing that fine-tuning mainly alters deeper encoder layers~\cite{elkheir2024, delaFuente2024}.

    \subsubsection{Models trained on general audio} \vspace{-1mm}
    \label{sec:analyse-env-sound}
    \begin{figure}[t]
        \centering
        \includegraphics[width=0.90\linewidth]{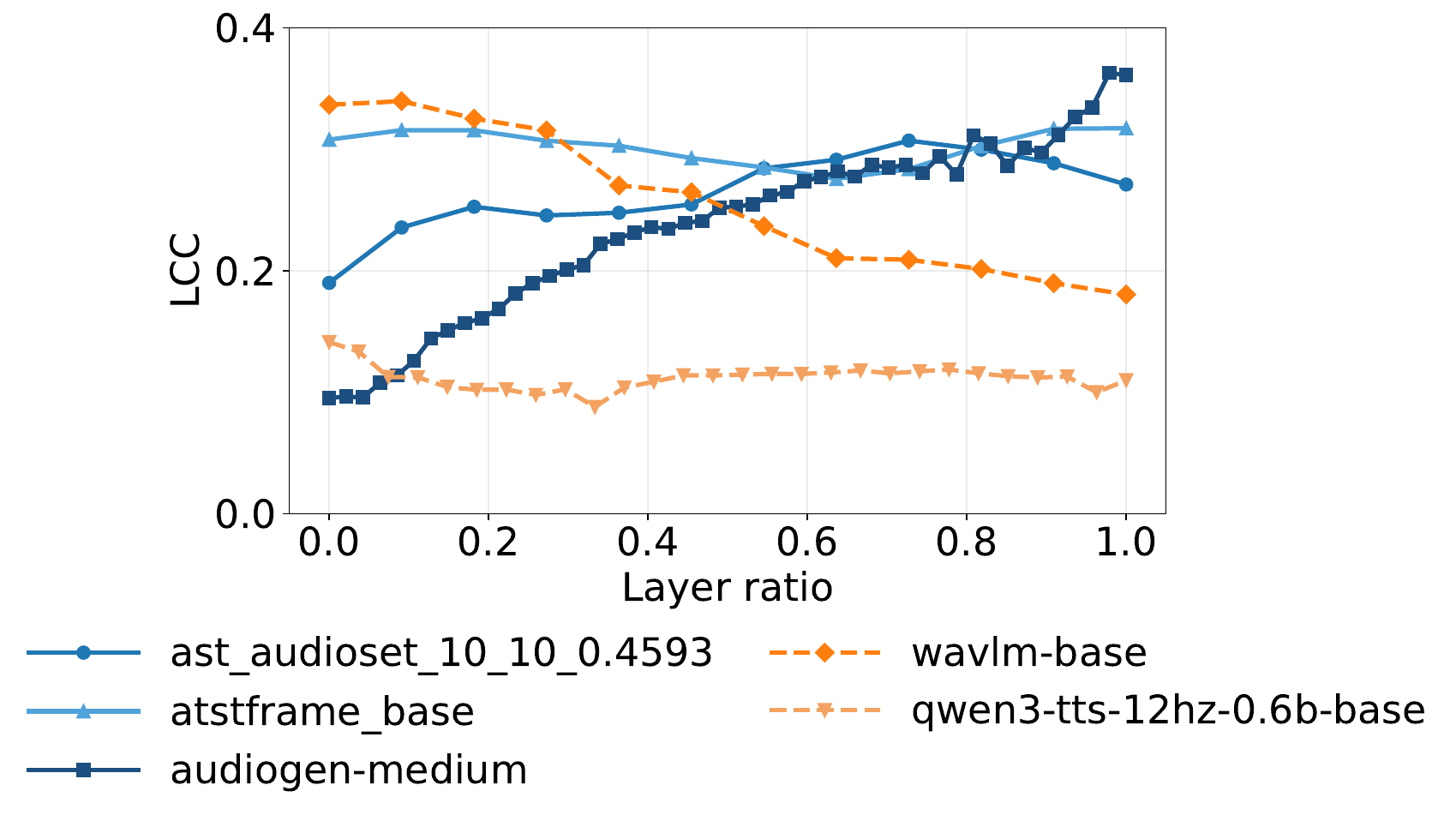}
        \vspace{-2mm}
        \caption{LCCs for models trained on general audio. For comparison, some models trained on speech are also drawn. }
        \vspace{-2mm}
        \label{fig:env_sound_lcc}
    \end{figure}
    
    To analyze how audio-related models relate to human perception, Fig.~\ref{fig:env_sound_lcc} shows LCC. Due to space limitations, SRCC, Frobenius distance, and spectral distance are omitted here, but they show the same trend as LCC.
    TTA and audio SSL models sometimes achieve higher scores than models trained solely on speech.
    Among the audio-related models, the TTA model \texttt{audiogen-medium} shows a monotonic increase as the layer depth increases. This behavior is likely because \texttt{audiogen} is trained as a sound generation model, where deeper layers represent more detailed timbral and acoustic characteristics, potentially leading to representations closer to perceptual speaker similarity.
    In contrast, \texttt{ast\_audioset\_10\_10\_0.4593}, which is trained with supervised learning for environmental sound classification, exhibits a peak in the middle layers. This may be because such categorical information is mainly encoded in intermediate representations; in later layers, the model likely recognizes human sounds primarily through event-level training labels such as speech, laughter, crying, or singing, rather than through fine-grained speaker characteristics.
    Finally, \texttt{atstframe\_base} shows a relatively flat trend. Since the model is trained on a broad range of sounds and is not specialized for any specific task, it is reasonable to treat audio representations uniformly across the entire network.


%% file: tables/mean.tex
\begin{table*}[t]
\caption{Coefficient ($p$-value) of multiple regression analysis using model configurations. \textbf{Bold} indicates coefficients with $p < 0.05$.}
\centering
\footnotesize
\begin{tabular}{l|rrrrr|r} 
Response variable & \texttt{is\_dec} & \texttt{is\_mlang} & \texttt{is\_ssl}                 & \texttt{hours} & \texttt{params} & $R^2$ \\ \midrule
\multicolumn{7}{l}{\texttt{layer\_max}}\\   \midrule
LCC & $\mathbf{-0.77}$ ($<.001$) &$0.00$ ($0.992$) &$\mathbf{-0.14}$ ($0.020$) &$-0.12$ ($0.117$) &$\mathbf{-0.14}$ ($0.022$) &$0.747$\\
SRCC & $\mathbf{-0.79}$ ($<.001$) &$-0.03$ ($0.687$) &$\mathbf{-0.15}$ ($0.015$) &$-0.10$ ($0.182$) &$\mathbf{-0.15}$ ($0.010$) &$0.764$\\
Neg. Frobenius norm&$\mathbf{-0.22}$ ($<.001$) &$-0.02$ ($0.407$) &$-0.01$ ($0.359$) &$-0.01$ ($0.568$) &$\mathbf{0.05}$ ($0.002$) & $0.983$\\
Neg. spectral dist.&$0.03$ ($0.529$) &$\mathbf{-0.16}$ ($0.008$) &$\mathbf{-0.15}$ ($0.003$) &$\mathbf{0.10}$ ($0.096$) &$-0.03$ ($0.550$) &$0.835$\\\midrule

\multicolumn{7}{l}{\texttt{layer\_slope}}\\   \midrule 
LCC&$\mathbf{0.31}$ ($0.002$) &$0.10$ ($0.425$) &$0.06$ ($0.610$) &$-0.26$ ($0.054$) &$\mathbf{0.29}$ ($0.008$) &$0.205$\\
SRCC&$\mathbf{0.29}$ ($0.004$) &$0.14$ ($0.297$) &$0.09$ ($0.424$) &$-0.24$ ($0.083$) &$\mathbf{0.26}$ ($0.017$) &$0.184$\\
Neg. Frobenius norm&$-0.12$ ($0.240$) &$0.20$ ($0.152$) &$-0.02$ ($0.890$) &$0.06$ ($0.650$) &$0.07$ ($0.553$) &$0.123$\\
Neg. spectral dist.&$0.09$ ($0.340$) &$\mathbf{0.54}$ ($<.001$) &$0.20$ ($0.058$) &$-0.15$ ($0.275$) &$0.13$ ($0.211$) &$0.241$
\end{tabular}
\label{tab:mean}
\end{table*}

%% file: sections/5.conclusion.tex
\section{Conclusion}
This study examined whether speaker representations in speech foundation models reflect human perceptual speaker similarity.
The results show that speech foundation models partially align with human perception, and that this alignment varies depending on model configuration.
We also observed that layer-wise behavior depends on model configuration and that fine-tuning objectives influence speaker representations. These findings provide insights into how model architecture and training strategies affect perceptually meaningful speaker representations.

%% file: sections/acknowledgments.tex
\section{Acknowledgments}
This work was supported by JST FOREST JPMJFR226V, JSPS KAKENHI 23K28108, and Moonshot R\&D Grant Number JPMJPS2011.

%% file: sections/disclosure_ai.tex
\section{Generative AI use disclosure}
ChatGPT was used for language polishing.